\documentclass[11pt]{article}
\usepackage{latexsym}
\usepackage{epsfig}
\usepackage{color}
\usepackage{verbatim}
\def\ed{\end{document}}

\def\beq{\begin{eqnarray}}
\def\eq{\end{eqnarray}}
\def\beqn{\begin{eqnarray*}}
\def\eqn{\end{eqnarray*}}

\def\nl{\noindent}

\begin{document}

\title {$Z^{\prime}_{B-L}$ phenomenology at LHC}
\author{ Y. A. Coutinho \\
Instituto de F\'isica--Universidade Federal do Rio de Janeiro\\
Av. Athos da Silveira Ramos 149 \\
 Rio de Janeiro - RJ, 21941-972, Brazil \\ \\
E. C. F. S. Fortes, J. C. Montero\\
 Instituto  de F\'\i
sica Te\'orica--Universidade Estadual Paulista \\
R. Dr. Bento Teobaldo Ferraz 271\\
S\~ao Paulo - SP, 01140-070, Brazil}
\date{}
\maketitle
\begin{abstract}
 We  study the $Z^\prime$ phenomenology for two extensions of the Electroweak Standard Model (SM) which have an extra $U(1)_{B-L}$ gauge factor.  We show the capabilities  of the LHC in distinguishing the signals coming from these two extensions and both of them from the Standard Model background. In order to compare the behavior of these $B-L$ models we consider the reaction $p + p\longrightarrow \mu^+ + \mu^- + X$ and compute some observables as the total cross sections, number of events, forward-backward asymmetry, final particle distributions like rapidity, transverse momentum, and dimuon invariant mass, for two LHC  regimes: $\sqrt{s}\,({\cal L})=7$ TeV  ($1\, \textrm{fb}^{-1}$ ) and $14 $ TeV ($100\, \textrm{fb}^{-1}$) for $M_{Z^{\prime}}$ = 1000 GeV and 1500 GeV. We show that by using  appropriate kinematic cuts some of the observables considered here are able to extract different properties of the $Z^\prime$  boson, and hence providing information about to which $B-L$ model it belongs to.
\end{abstract}

\noindent

{elaine@ift.unesp.br; montero@ift.unesp.br;yara@if.ufrj.br}

\vfill\eject

\section{Introduction}
\label{sec:intro}

 The starting up of the Large Hadron Collider (LHC) operating in an energy range far above the electroweak scale, offers a  possibility to reveal new phenomena and to explore the  phenomenology extracted from the expected huge amount of experimental data. In this way, it is expected new physics manifestations throughout the appearance of new
degrees of freedom such as new charged and neutral fermions, superpartners, new gauge bosons, and  Higgs scalar(s). To take into account these new possible degrees of freedom we must go beyond the SM. There are many ways to construct extensions of the electroweak SM. However, if we are concerned with new gauge bosons we must consider a larger gauge group. Whatever the extension of the SM is, it must present the same results of the SM at the electroweak energy scale in order to be consistent with present data.

Although hadron colliders present a high background (pile-up)
compared to electron-positron (beamstrahlung) colliders,
they seem to be more suitable for discovering new degrees of freedom.
Since high centre-of-mass energy and high precision can not be  achieved
simultaneously in the same experiment, both types of colliders are needed in order to have a detailed description of a particular phenomena.
This is easily realized if we look to a recent past, when the synergy between hadron and
lepton colliders, brought up complementary information for each
other, generating benefits to high energy physics. As examples we
have the discovery of the $Z$ boson, the gluon, and the quark top. The
discovery of the $Z$ boson occurred in a $p \bar p$ collider (SPS) but its
properties (decay widths, couplings to fermions, asymmetries,
mass, etc...) were measured with high precision in lepton
colliders (SLC, LEP). In this way, the
recent studies started at LHC can direct the search for new
physics in future linear colliders (ILC/CLIC).

Among the main goals of any accelerator proposal  we find
the physics of an extra neutral gauge boson $Z^{\prime}$. This new boson is  predicted by many
extensions of the  Standard Model like:

\begin{itemize}
\item The models from $E_{6}$ group, also know as rank 5
  models (ER5M) \cite{lrgut, e6};
\item The Left-Right which are based in the group $SU(2)_{L} \otimes SU(2)_{R} \otimes
U(1)_{B-L}$ \cite{lr};
\item The models with  $Z^{\prime}$ in Little Higgs  scenario \cite{Lit};
\item The Sequential Standard  models - SSM \cite{Ssm};
\item The Kaluza Klein (KK) models, predicted by extra dimensions \cite{Riz};
 \item  The models with $SU(3)_c\times SU(3)_L\times U(1)_{N}$ gauge symmetry, known as  3-3-1 models \cite{PIV, FRA, TON};
\item Models where a strong dynamics is involved in the electroweak  spontaneous symmetry breaking. We can cite, for instance, Topcolor and BESS (Breaking Electroweak Symmetry Strongly) models \cite{Bess}.
\end{itemize}

Some models which have an extra $U(1)$ gauge group~\cite{Appel, Blsm} also deserve  a detailed study  in order to predict new physics behavior. The cumulative study  of different models, candidates to describe the electroweak interaction,  can guide the LHC data analysis if in fact a $Z^{\prime}$ is discovered. The extra $U(1)$ group gives rise to the $Z^{\prime}$ vector boson. The new constraints on these models generate few free parameters, if we compare them  to other models which have  origin in larger gauge groups.  Following the arguments here exposed, first we have  considered  the $Z^{\prime}$ boson observables for electron-positron colliders \cite{Elai}, and now we extend  the study for hadron colliders.

In this paper, we study the phenomenology of the $Z^{ \prime}$ boson originated from two different extensions of the electroweak SM with an extra $U(1)$ gauge factor.
We predict the possible signals of new physics in the process $p + p \longrightarrow  (\gamma, \,\, Z, \,\, Z^{
\prime})\longrightarrow \mu^+ + \mu^- + X$, at the tree level.
We compare several quantities  as total
cross sections, number of events, and decay widths assuming that both $U(1)$ factors correspond to the local $B-L$ symmetry, and that the breaking of  this symmetry occurs at an energy scale above the TeV scale.

In order to have a more realistic scenario, we
choose the parameters and constraints of the two different models to compare them at the same $Z^\prime$ boson mass. We use $M_{Z^{\prime}}= 1500$ GeV, but we also show the results for $M_{Z^{\prime}}= 1000$ GeV, for two centre-of-mass energies: $\sqrt{s}=7$ and $14$ TeV.

This article is organized in the following way. In section II we introduce the models. In section III we present the results and in section IV we present our conclusions.

\section{The models}
\label{sec:models}

  In this paper we consider two extensions of the electroweak SM  which have an extra $U(1)$ local factor, resulting in the total gauge symmetry $ SU(2)_L\otimes U(1)_1\otimes U(1)_2$. More specifically, we are concerned with models which are based on the gauge symmetries:

\begin{itemize}
\item $SU(2)_L\otimes U(1)_Y\otimes U(1)_z$, called as Secluded model ~\cite{Appel};
\item $SU(2)_L\otimes U(1)_{Y^\prime}\otimes U(1)_{_{B-L}}$  called as Flipped model ~\cite{Blsm}.
\end{itemize}

In the Secluded model, $z$ is a new $U(1)$ charge, and $Y$ is the electroweak SM hypercharge. In the Flipped one, $Y^\prime$ is a new extra $U(1)$ charge, and the $B-L$ charge assignments are those of the SM.
In this last model the SM  hypercharge is recovered after the first spontaneous symmetry breaking.

In the Secluded model, the electric charge has no component in the $U(1)_{z}$ and in particular, depending on the value of this charge, several versions of this model can be studied. Our analysis will be focused on comparing both models only when the $z$ charge of the Secluded model coincides with $B-L$. Besides, in this minimal version, considered here, both models have the same scalar sector, composed by one doublet, $H$, with  hypercharge $Y=+1$, and a complex singlet with $Y=0$.  The number of right-handed neutrinos is three, in order to be consistent with the anomaly  cancellations. We had set them to be heavy enough in order to have $Z^{\prime}$ decay into heavy right-handed neutrinos kinematically suppressed. Besides that, the minimal version considered here is also absent of exotic decay channels like heavy quarks or heavy charged leptons. So, our results of decay widths, branchings fractions for these two particular $Z^{\prime}$ models are related only to standard model fermions. We will parameterize the neutral currents in terms of the mass eigenstate fields as follows:
\begin{equation}
{\mathcal{ L}}^{NC}=- \frac{g}{2c_W}\sum_i  \overline{ \psi_i}
\gamma_\mu [(g^i_V-g^i_A \gamma_5 ) Z^\mu_1 + (f^i_V-
f^i_A\gamma_5 )Z^\mu_2]\psi_i.
\label{nc}
\end{equation}

\par
 In this paper we briefly present both models. The complete details about both of them were subject of study in a recent work \cite{Elai}. Moreover, the scalar sector of this sort of models, with the respective Feynman rules, is considered in reference \cite{Khalil}.
The charge  operators for the Secluded and  the Flipped models, are respectively given by:
    \begin{eqnarray*}
    \frac{Q}{e}=I_3 + \frac{1}{2}\,Y,
    \label{gn2}
\end{eqnarray*}
\begin{eqnarray*}
\frac{Q}{e}=I_3+\frac{1}{2}\,\left[Y^\prime + (B-L)\right].
    \label{gn1}\end{eqnarray*}

When the $z$ charge of the Secluded model is equal to $B-L$ it implies  that the parameter $z_{H}$ (the $z$-charge of the 
SM Higgs doublet), discussed in Refs.~\cite{Appel,Blsm}, is equal to zero. In this case, there is no tree level mixing in 
the mass matrix  between the neutral gauge bosons $B_{z}^{\mu}$, $W_{3}^{\mu}$ and $B_{Y}^{\mu}$, and we have $Z=Z_{1}$ and 
$Z^{\prime}=Z_{2}$, i.e. the mass eigenstates and the symmetry eigenstates coincide. This is not the case for the  Flipped 
model. Here, we will assume that the mixing between the two neutral vector bosons is  small enough so that $Z \approx Z_{1}$
and $Z' \approx Z_{2}$.

Both models are well motivated since, in general, they can be thought as an intermediate symmetry, remnant of a larger 
(unknown) gauge symmetry, that was broken at a very large energy scale.  Moreover, they are simpler compared to other class 
of models with an extra $Z^\prime$ neutral vector boson: simpler gauge groups and fewer free parameters.

\section{Results}

We have first chosen $M_{Z^{\prime}}= 1500$ GeV in order to compare the results coming from the two different models.  For the Secluded model, this choice respects the bounds already established on the $Z^{\prime}$ boson mass and its coupling constants by the direct search at the Tevatron~\cite{Abu},
by the electroweak precision tests (EWPT) at LEP II, and by the low-energy neutral current experiments \cite{Erler, Caccia, Salvioni} , which are well
studied in the literature.  On the other hand, the Flipped model has not received the same level of attention. We find that it also deserves a detailed  study to establish constraints on its parameters and numerical couplings.

For the Flipped  model we use the following inputs: $g_{_{B-L}}=0.6132$, $g^\prime=0.4400$, $u= 1987$ GeV, $v = 246$ GeV,
$\alpha=127.9$, and $s^2_W=0.23122$ \cite{Pdg}. With these parameters  we
obtain $M_{Z^\prime}= 1500$ GeV, $\Gamma_{Z^{\prime}}= 28.39$ GeV, and the values of the neutral current coupling constants 
shown in Table 1. The $Z$ couplings with SM fermions remain the same so we emphasize only the new contribution from the 
$Z^{\prime}$ boson.

\begin{table}\label{Acoplflipp}
\begin{eqnarray*}
\begin{array}{|c||c|c|c|c|}
  \hline
 &Neutrinos & Leptons & u-quarks & d-quarks \\ \hline\hline
  f_V& 0.8420 & 0.4977 & -0.0511 & -0.3955  \\ \hline
 f_A& -0.1732 & 0.1715 & -0.1732 & 0.1732 \\ \hline
\end{array}
\end{eqnarray*}
\caption{$Z^\prime$ coupling constants to the SM fermions for the Flipped model for $M_{Z^{\prime}}= 1500$ GeV.}
\end{table}

For the Secluded model we use as inputs: $z_H=0$, $g_z=0.2$,
$z_q = z_u=1/3$, $u=7500$ GeV, and $v=246$ GeV. With these
parameters we obtain $M_{Z^\prime}=1500$ GeV,
$\Gamma_{Z^{\prime}}= 4.04$ GeV, and the values of the neutral
current coupling constants shown in Table 2. Due to the choice
$z_H=0$ in the Secluded model, the
$Z^\prime$  boson has only vector couplings to fermions, and presents a \textit{leptofilic} behavior around the peak region. The \textit{leptofilic}   character means that the $Z^\prime$ boson couples preferentially to leptons, and that the cross section for lepton production presents a higher peak compared to  quark cross sections. This behavior can be seen with more evidence  in $e^+ + e^-
\longrightarrow l + \bar l$ process, where $l$ is a lepton.

\begin{table}\label{Acoplsecl}
\begin{eqnarray*}
\begin{array}{|c||c|c|c|c|}
  \hline
 & Neutrinos & Leptons & u-quarks & d-quarks \\ \hline\hline
  f_V & 0.2690 & 0.2690 & -0.0897 & -0.0897  \\ \hline
 f_A & 0 & 0 & 0 & 0 \\ \hline
\end{array}
\end{eqnarray*}
\caption{$Z^\prime$ coupling constants to the SM fermions for the Secluded model for $M_{Z^{\prime}}= 1500$ GeV.}
\end{table}

We can realize through the partial decay widths that, as we  said before, for the Secluded model, $Z^{\prime}$
couples preferentially to leptons, as  shown in  Table 3. Due
to this character, and depending on the choice of the $Z^\prime$
parameters,  the Secluded model can give an
explanation for the positron excess in cosmic rays, as presented by the PAMELA
experiment~\cite{Pam}.

\begin{table}\label{Branching}
\begin{footnotesize}
\begin{center}
\begin{tabular}{||c|c|c||}
     \hline
$M_{Z^{\prime}}= 1500$ GeV &  Flipped $B-L$ &  Secluded $B-L$ \\
\hline
$Z^{\prime} \rightarrow \sum_i \, \bar \nu_i \nu_i$ & $36\%$ &
$37.5\%$  \\
\hline
$Z^{\prime} \rightarrow \sum_i \,\bar l_i l_i$ & $18.6\%$
& $37.5\%$  \\
\hline
$Z^{\prime} \rightarrow \sum_i \,\bar q_i q_i$  &   $42.6\%$  &
$25\%$ \\
\hline
$Z^{\prime}  \rightarrow W^+W^-$  &   $2.8\%$  &
$0\%$  \\
\hline
\end{tabular}
\end{center}
\end{footnotesize}
\caption{The $Z^{\prime}$ partial decay widths for fermions and charged vector boson for both $B-L$ models.}
\end{table}

 We present in Figure 1 the total $Z^\prime$ decay width against
$M_{Z^\prime}$. Both of them are linear functions of $M_{Z^\prime}$, as it should be, and hence they double their 
values in the $Z^\prime$ boson mass range showed in the figure. However, according to the parameter choice, $g_{z}=0.2$, 
the decay width of the Secluded model is smaller than that of the Flipped one. In the latter model the $Z^\prime$ boson  
has both axial and vector couplings, and in the Secluded model it has only the vector ones due to the choice $z_{H}=0$.

\begin{figure}[ht]
\begin{center}
\includegraphics[height=.3\textheight]{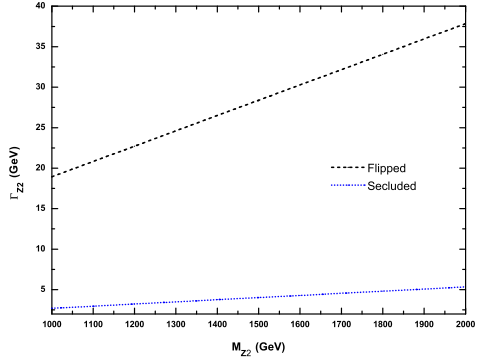}
\end{center}
\caption{\label{fig15} Evolution of total decay width with $Z^\prime$ boson mass.}
\end{figure}

The ${Z^{\prime}}$ production can be achieved by the Drell-Yan
mechanism which implies in leptonic decay products accompanied by
background, which contaminates the observables under consideration. This
background can be minimized applying some specific cuts. To
perform our calculations we use the package Comphep \cite{Comp}
with  Cteq6L1 parton distribution functions. For the
first simulations we have used the following cuts on the final
leptons, which we call as the first cut set: $p_{t \mu} > 20 $
GeV, $\vert \eta_{\mu} \vert
\le 2.5 $ and $ -0.99 <  \cos\theta_{q\mu}  < 0.99 $.
\\
The first observable obtained, when the first cut set is used, is
the total cross section for $ p + p\,\,\,
{\rightarrow{(\gamma, Z, Z^{ \prime})}} \rightarrow\,\,\,\mu^+ + \mu^-
+ X $. Considering $M_{Z^\prime}= 1500$ GeV we obtain for the Flipped, Secluded and SM,
 $ 1.33\times 10^3 $ pb, $ 1.01\times 10^3 $ pb,  and
$ 0.6\times 10^3 $ pb, respectively for $\sqrt s = 14 $ TeV, which implies
considering the annual luminosity $ {\cal L} = 100 \,\,
{\hbox{fb}}^{-1} $ in a huge number of events: $ \sim 10^8 $. We present in Table 4 the number of events considering  $M_{Z^\prime}= 1500$ and 1000 GeV for both models and also  $\sqrt{s}=14$ and 7 TeV. The number of
events will decrease when we apply more restrictive cuts. In order to assign the correct quark direction,
we had selected dimuon large rapidity events, $\vert y_{\mu \mu} \vert > 0.8$, as suggested in \cite{Dit, Nic}, the other cuts
are $ p_{t\mu} > 20 \,\,\, {\hbox{GeV}}$,  $ \vert \eta_{\mu} \vert \le 2.5 $, $ 1450
\,\,\, {\hbox{GeV}} < M_{\mu \mu} < 1550 \,\,\, {\hbox{GeV}} $, and
$- 0.99 < \cos\theta_{q\mu} < 0.99 $, and we call them as the second cut set.

\begin{table}\label{Branching1}
\begin{footnotesize}
\begin{center}
\begin{tabular}{||c|c|c|c||}
     \hline
Model & $M_{Z^{\prime}}= 1500$ GeV  & $M_{Z^{\prime}}= 1000$ GeV & $M_{Z^{\prime}}= 1000$ GeV \\
\hline
Flipped $B-L$ & $1.33\times 10^{8}$ & $8.48\times 10^{3}$ & $245$\\
\hline
Secluded $B-L$ & $1.01\times 10^{8}$ & $ 7.18\times 10^{3}$ & $219$\\
\hline
\end{tabular}
\end{center}
\end{footnotesize}
\caption {Number of events considering $M_{Z^{\prime}}= 1500$ and  1000 GeV, and the first cut set. The second and third 
columns are  for $\sqrt s = 14$ and $\mathcal{L}= 100 fb^{-1}$, and the forth column is for $\sqrt s =7$ TeV 
and $\mathcal{L}= 1 fb^{-1}$. In the data related to $M_{Z^{\prime}}= 1000$ GeV we have also set  $M_{\mu\mu}>500$ GeV.}
\end{table}

Besides the total cross section and decay width for leptonic final
state, we present: invariant mass, rapidity,
forward-backward asymmetry, transverse momentum and angular
distribution. In Figures 2a and 2b we present the dilepton
invariant mass distribution. The Flipped model presents a distinct behavior with a higher peak far above the Standard model and has very good changes to be separated from the background. On the other, the Secluded model
distribution in the region off--peak in Figure 2 (down panel) has a  behavior similar to that of
the SM, while the on--peak region is not so powerful to
distinguish it from the SM. This difference has origin mainly in
the choice of $z_H = 0 $ for the Secluded model, which leads to
pure vector couplings  of the ${Z^{\prime}}$ boson to fermions. A different  choice for $z_H$, say
 $z_H \neq 0 $,  will restore the axial couplings but, in this case, the $z$-charge will not be equal to $B-L$.
 The case of $z_H \neq 0 $ is out of the scope of the present work.

\begin{figure}[ht]
\begin{center}
\includegraphics[height=.3\textheight]{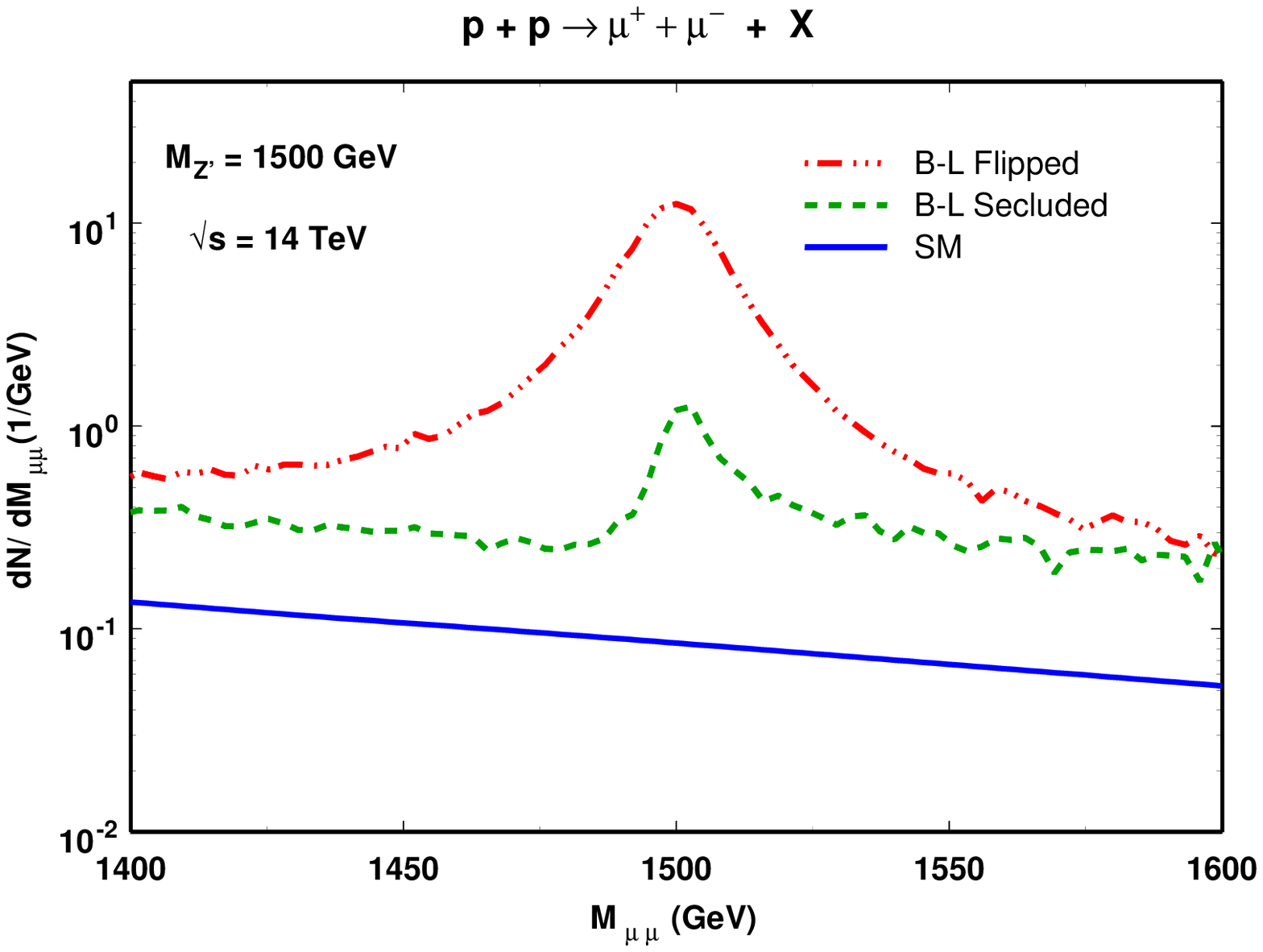}
\includegraphics[height=.3\textheight]{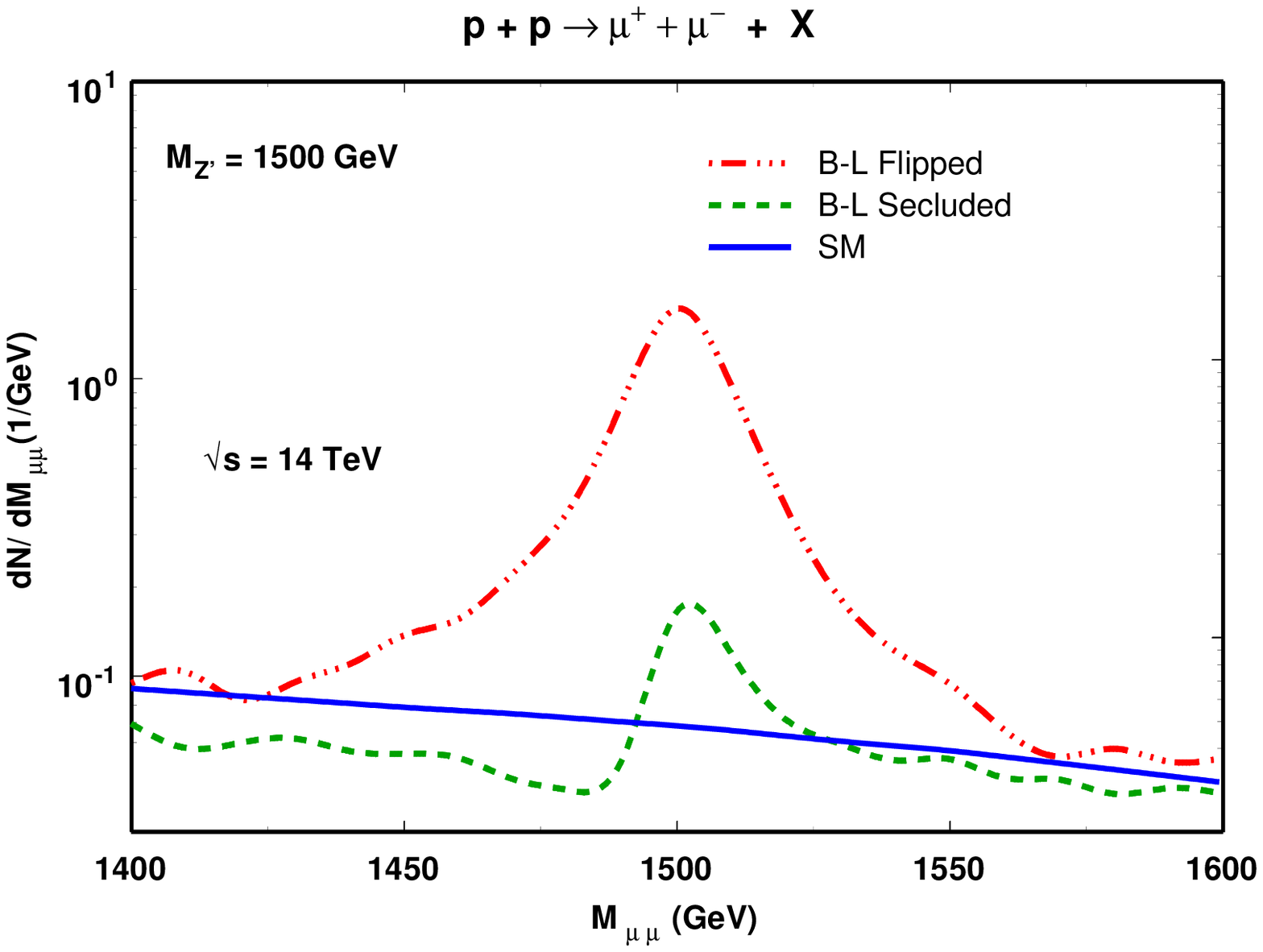}
\end{center}
\caption{\label{fig3} The dilepton invariant mass distribution for $ p
+ p \to \mu^{+} + \mu^{-} + X$ process for the Flipped, Secluded and
Standard models considering $M_{Z^{\prime}}= 1500$ GeV at $\sqrt s=14 $ TeV. First
cut set (up) and second cut set (down).}
\end{figure}

The rapidity distributions can also be used to  disentangle the $B-L$ models.
This observable is powerful even when we consider both
light and heavy quarks or only the light ones, this result is
showed in Figures 3a, 3b and 3c. The signal has been improved for both
models but once again the Flipped model produces a signal above the ones of
the other two models. One intrinsic characteristic of the Flipped
model is the increasing signal until it reaches the region where the rapidity of each muon is $\vert y_{ \mu}\vert =
1.25$. Both, the Secluded model and the SM do not have this behavior. When we
apply more severe cuts on rapidity, as $\vert y_{\mu} \vert >
0.8$, the resulting data reveal that the three models have less possibilities
to be disentangled. So, for this observable, the first cut set is
more efficient than the more severe ones.

\begin{figure}[ht]
\begin{center}
\includegraphics[height=.3\textheight]{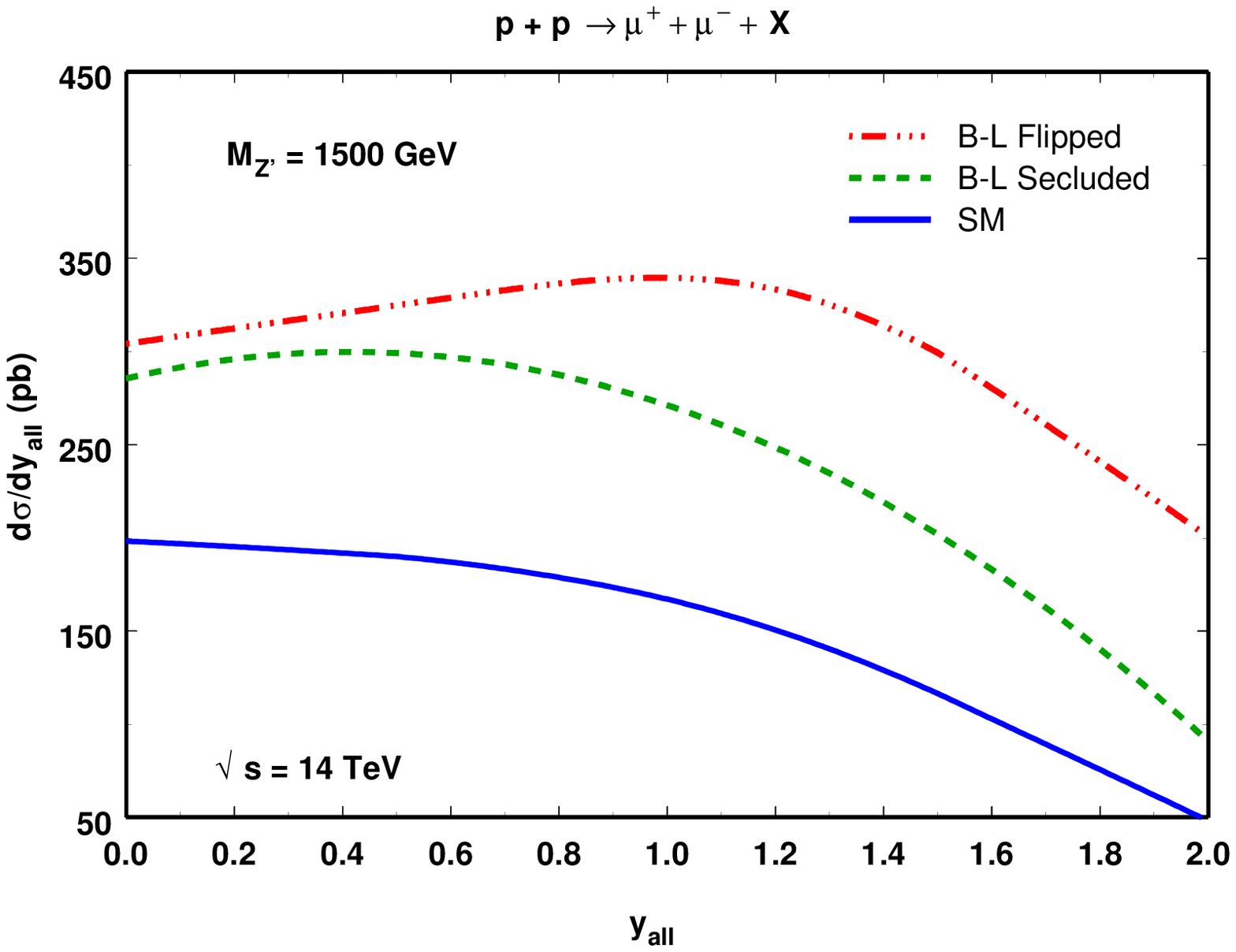}
\vglue -0.8cm
\includegraphics[height=.3\textheight]{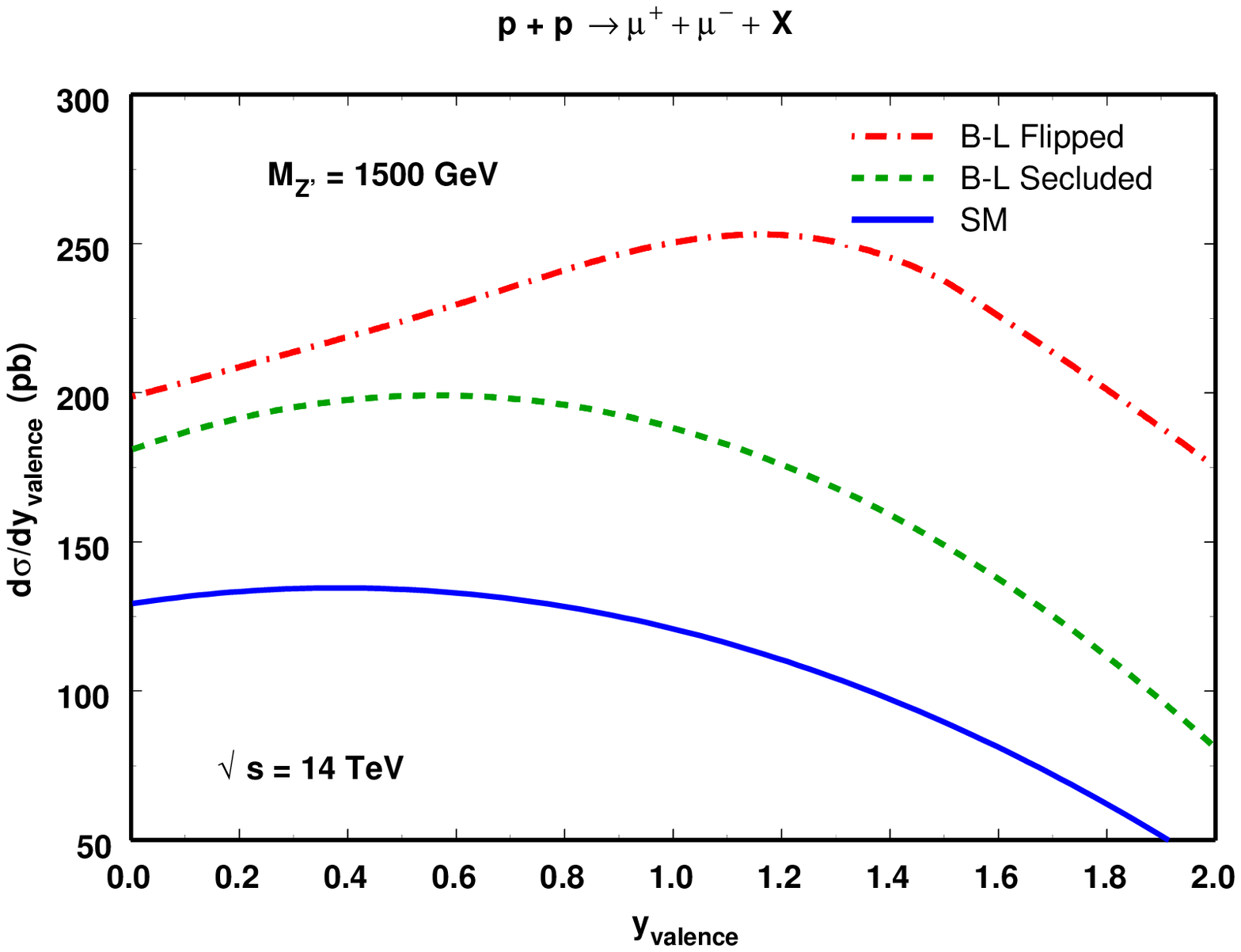}
\vglue -0.8cm
\includegraphics[height=.3\textheight]{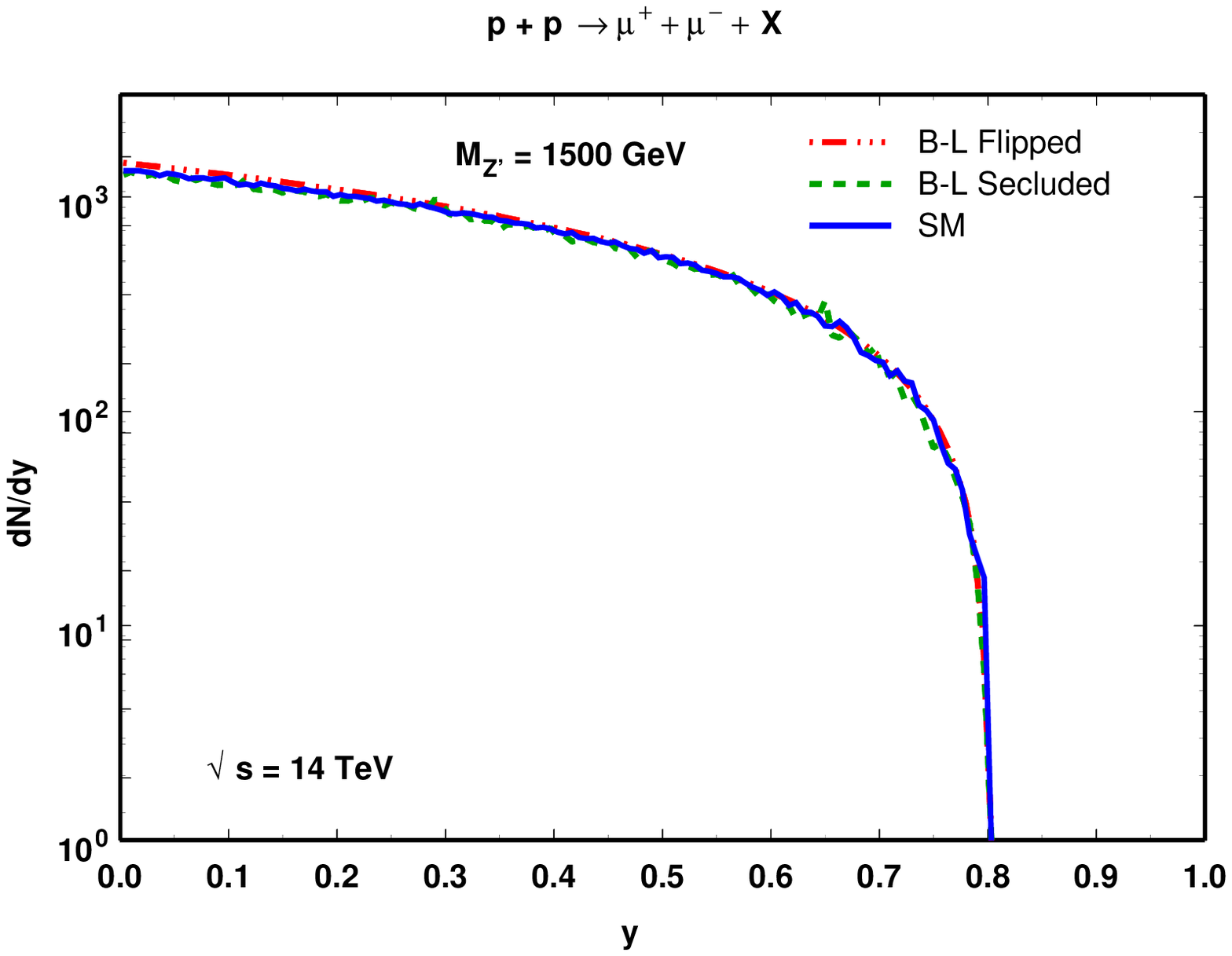}
\vglue -1.5cm
\end{center}
\caption{The rapidity distribution for the process $ p + p \to \mu^{+}
+ \mu^{-} + X$ considering $M_{Z^{\prime}}= 1500$ GeV  for the Flipped,
Secluded and Standard models at  $\sqrt s = 14$ TeV. The valence and sea quark contributions (up) and only
the valence quarks (middle), using the first cut set. The same
rapidity distribution using the second cut set (down).}
\end{figure}

Another important observable, very sensitive to the new physics
contributions, is the forward-backward asymmetry ($A_{FB}^{\ell}$).
In  $e^+ e^-$ colliders,  $A_{FB}^{\ell}$ is
measured with high precision, due to the well known initial beam
direction.  However, the situation is quite different for hadron colliders where the original quark direction is completely unknown. To solve this problem, the quark direction can
be approximated by the boost direction, which connects the dimuon
reference system with the original beam direction. In order to achieve this
goal we have applied the $\vert y_{\mu \mu}
\vert > 0.8$ cut  for the muon pair rapidity,  following the references \cite{Dit, Nic}.
 In contrast with the Flipped model, the Secluded one has no $ Z^{\prime}$
axial couplings to fermions and, as a consequence of it, the forward-backward
asymmetry, even being a powerful tool to get information on the
$Z^{\prime}$ boson parameters, does not receive contributions from the axial
couplings on-peak. In the Figure 4 we show this observable
versus $M_{\mu \mu}$, and we see that  the two $B-L$ models can be
distinguished.

\begin{figure}[ht]
\begin{center}
\includegraphics[height=.3\textheight]{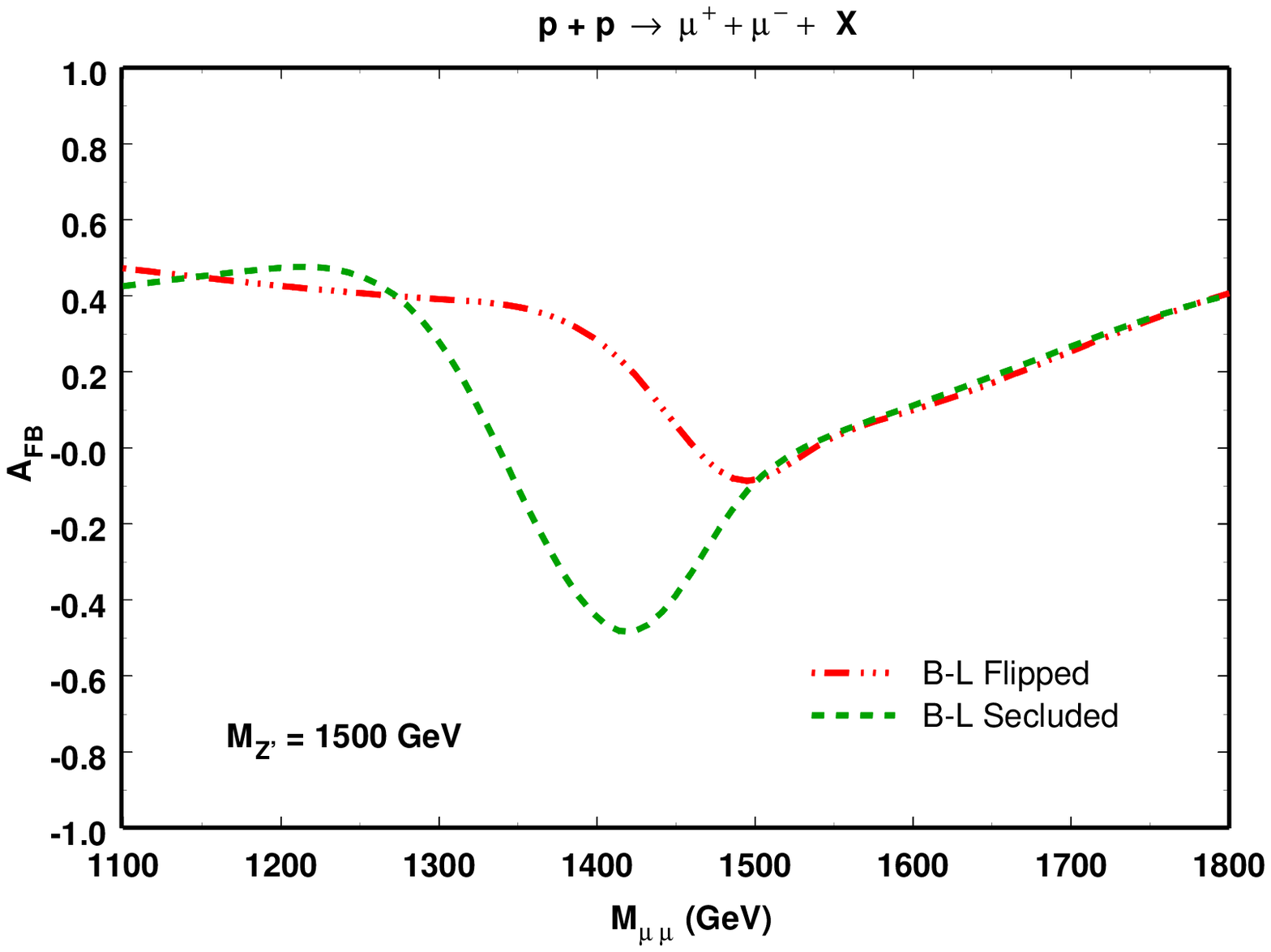}
\end{center}
\caption{The forward-backward asymmetry  for the process $ p + p
\to \mu^{+} + \mu^{-} + X$  as a function of the dilepton invariant
mass for both $B-L$ models, considering $M_{Z^{\prime}}= 1500$ GeV, $\sqrt s = 14$
TeV, and applying the second cut set.}
\end{figure}

If we consider all the studied observables, after applying the second cut set, we will realize that one particular numerical cut can
put in evidence one observable and  others not.  The cut $\vert
y_{\mu \mu} \vert
> 0.8$ is not so helpful for rapidity distributions, but it is
necessary for forward-backward asymmetries in order to guarantee
 that the quark direction has more probability to be the correct one. It means that the
probability of obtaining more significant measurements of forward-backward asymmetries
increases for higher rapidities.

We present in the Figure 5 our results for the muon transverse
momentum distribution.  One of the characteristics of this
distribution is that the peak is located right at the position
$p_T = M_{Z^\prime}/2$, for both $B-L$ models. We have adopted more restricted cuts in the final dimuon mass,  $M_{\mu\, \mu}
> 500$ GeV \cite{Elm} in order to emphasize this behavior. The flipped has once again, much better chances to be separated 
from the background, in opposition, the Secluded one can be mistaken as Standard model background.

\begin{figure}[ht]
\begin{center}
\includegraphics[height=.3\textheight]{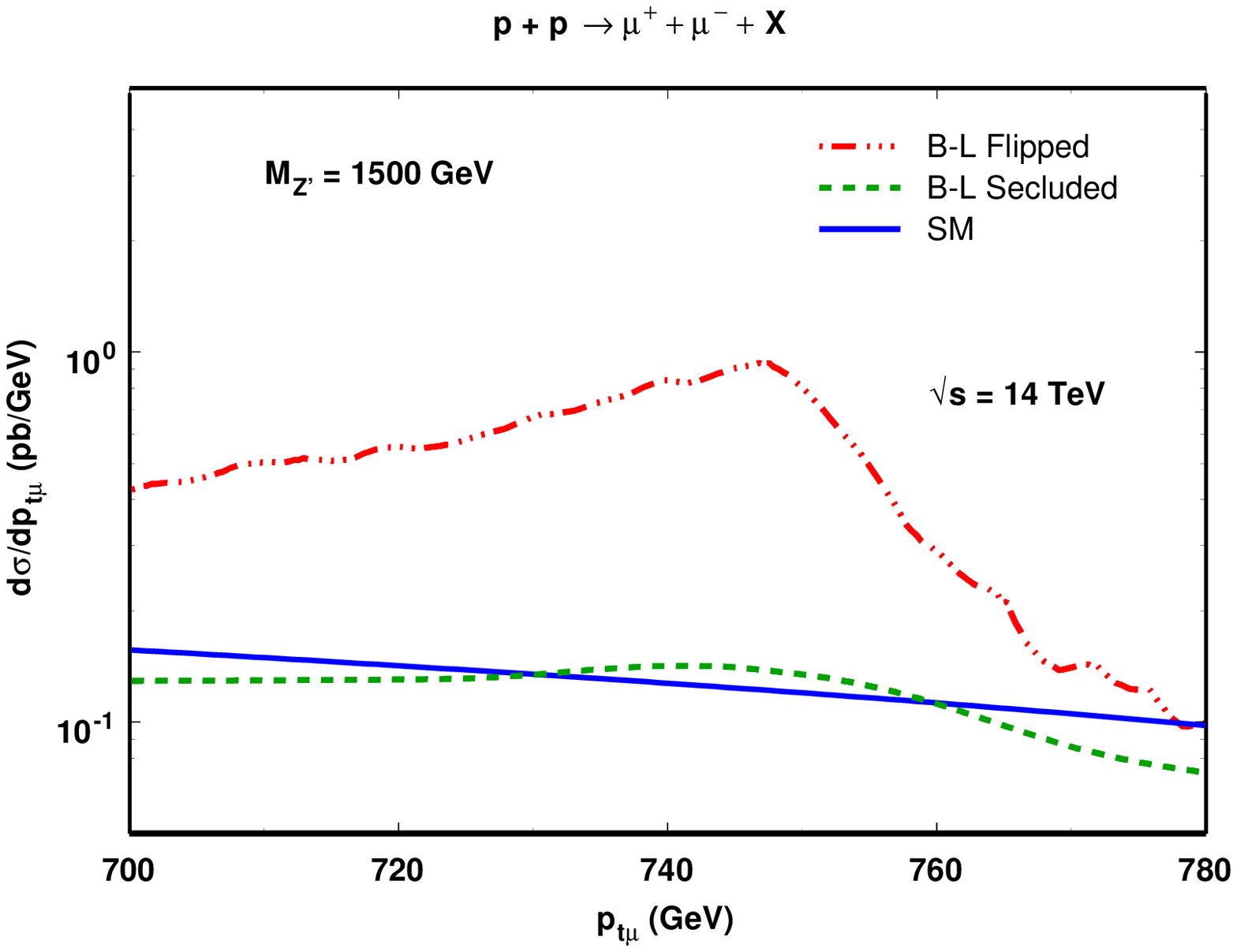}
\end{center}
\caption{The muon transverse momentum distribution for the process
$ p + p \to \mu^{+} + \mu^{-} + X$ for the Flipped, Secluded and Standard models, considering $M_{Z^{\prime}}= 1500$ GeV, $\sqrt s = 14$ TeV, and applying the second cut set.}
\end{figure}

The muon angular distribution is less sensitive to ${Z^{\prime}}$
contributions, but can be  used as a previous result before the
calculation of $A_{FB}^{\ell}$. From Figure 6 we can realize
that the signals related to the SM are below the signals of the two
$B-L$  models. We can see a small asymmetry when the
curves cross each other in the region $-0.50 < \cos \theta <
-0.25$, where  $\theta$ is the angle between the beam and the muon directions.

\begin{figure}[ht]
\begin{center}
\includegraphics[height=.3\textheight]{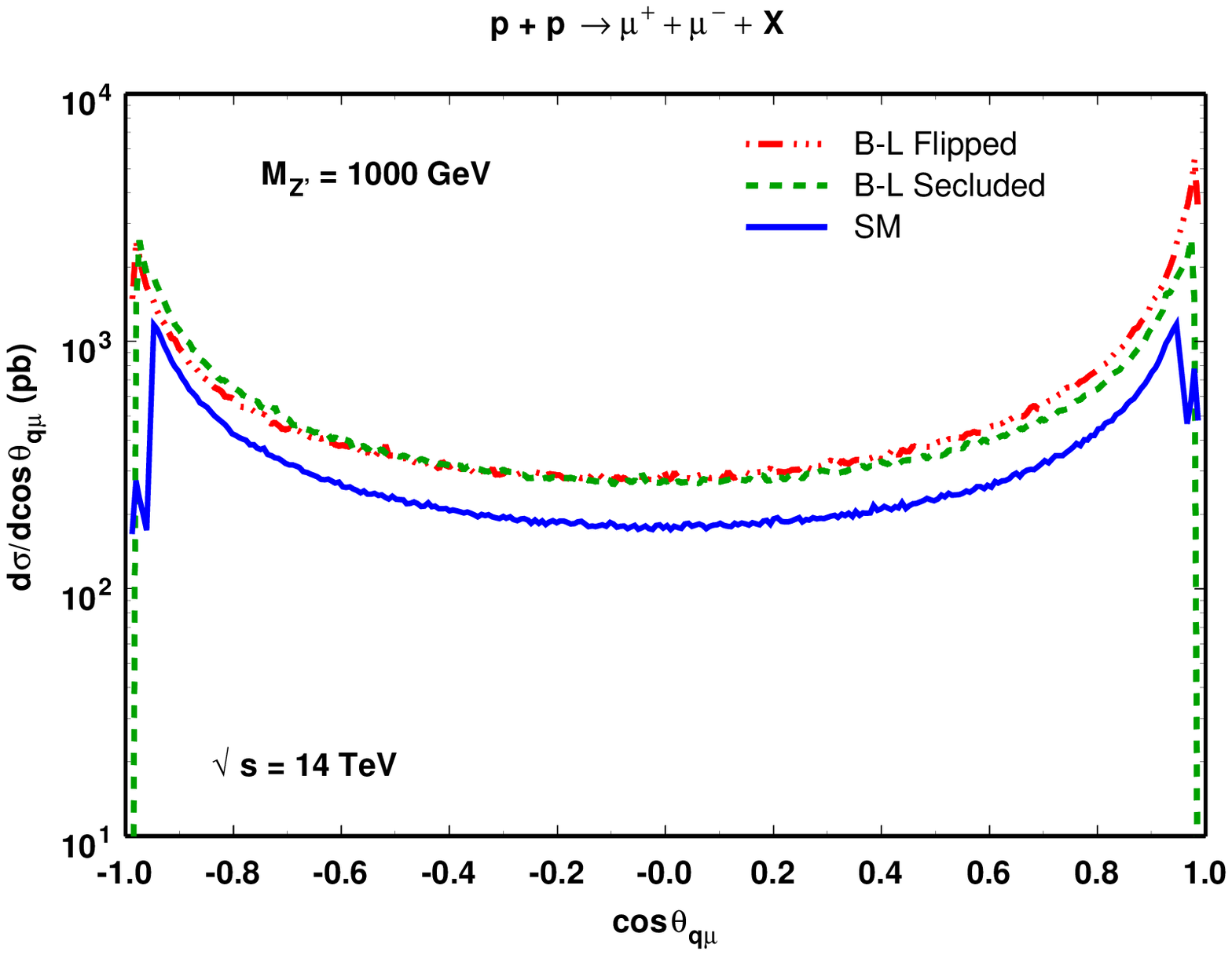}
\end{center}
\caption{The muon angular distribution for the process
$ p + p \to \mu^{+} + \mu^{-} + X$ for the Flipped, Secluded and Standard models,  considering $M_{Z^{\prime}}= 1000$ GeV and $\sqrt s = 14$ TeV.}
\end{figure}

We now consider  a ${Z^{\prime}}$ boson which mass is 1000 GeV and  $\sqrt{s}=7$ and $14$  TeV in order to
 extract more conclusions about the same observables presented above. The related graphs  are presented in the  
Figures 7, 8, 9 and 10. The increase of the centre of mass energy from 7 to 14 TeV has as direct consequences the 
increase of the cross sections and, hence, the number of events. The invariant mass distribution shows also a small peak for the  Secluded model if compared to the  Flipped one, as we can see in  Figures 7a and 7b. The  rapidity distributions
 presented in Figures 8a and 8b show that for $M_{Z^{\prime}}= 1000$ GeV  both models can not be clearly distinguished 
from each other in this case. If we consider $\sqrt{s}=14$ TeV the rapidity distributions graphs are separated, mainly in the 
region  $0 < y_{\mu \mu} < 1$, but this separation is not so noticeable as the one for $M_{Z^\prime}= 1500$ GeV, already 
presented in Figures 3a and 3b. The transverse  momentum distributions, presented in the  Figures 9a and 9b, do not show a 
clear peak for $M_{Z^\prime}/2$ for the  Secluded model, but the Flipped one is put in evidence once again with higher values 
for this distribution on--peak.  The muon angular distributions  present a better behavior for disentangling both models, 
mainly if we consider $\sqrt{s}=14$ TeV, as  can be seen from Figures 10a and 10b.

\begin{figure}[ht]
\begin{center}
\includegraphics[height=.3\textheight]{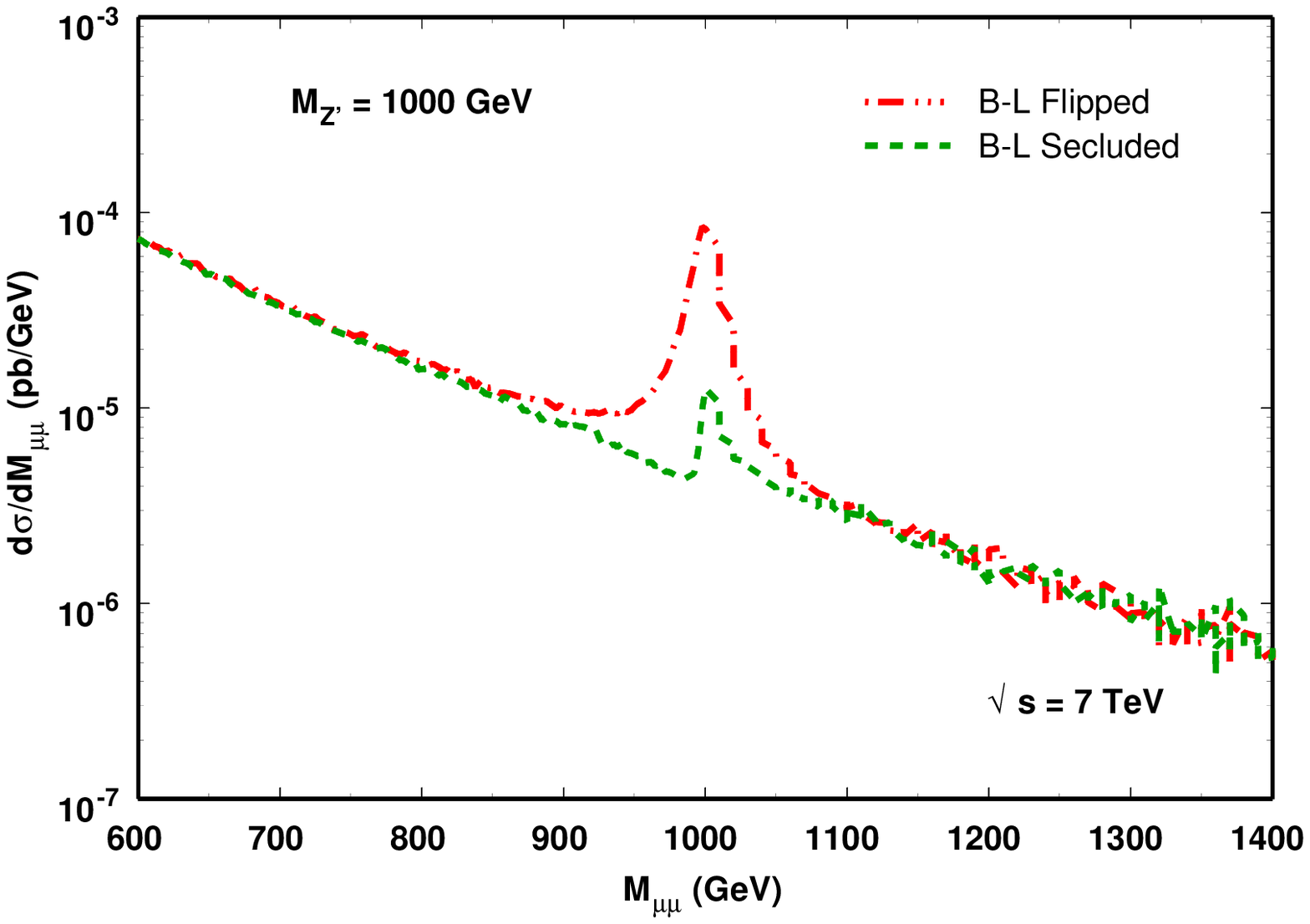}
\includegraphics[height=.3\textheight]{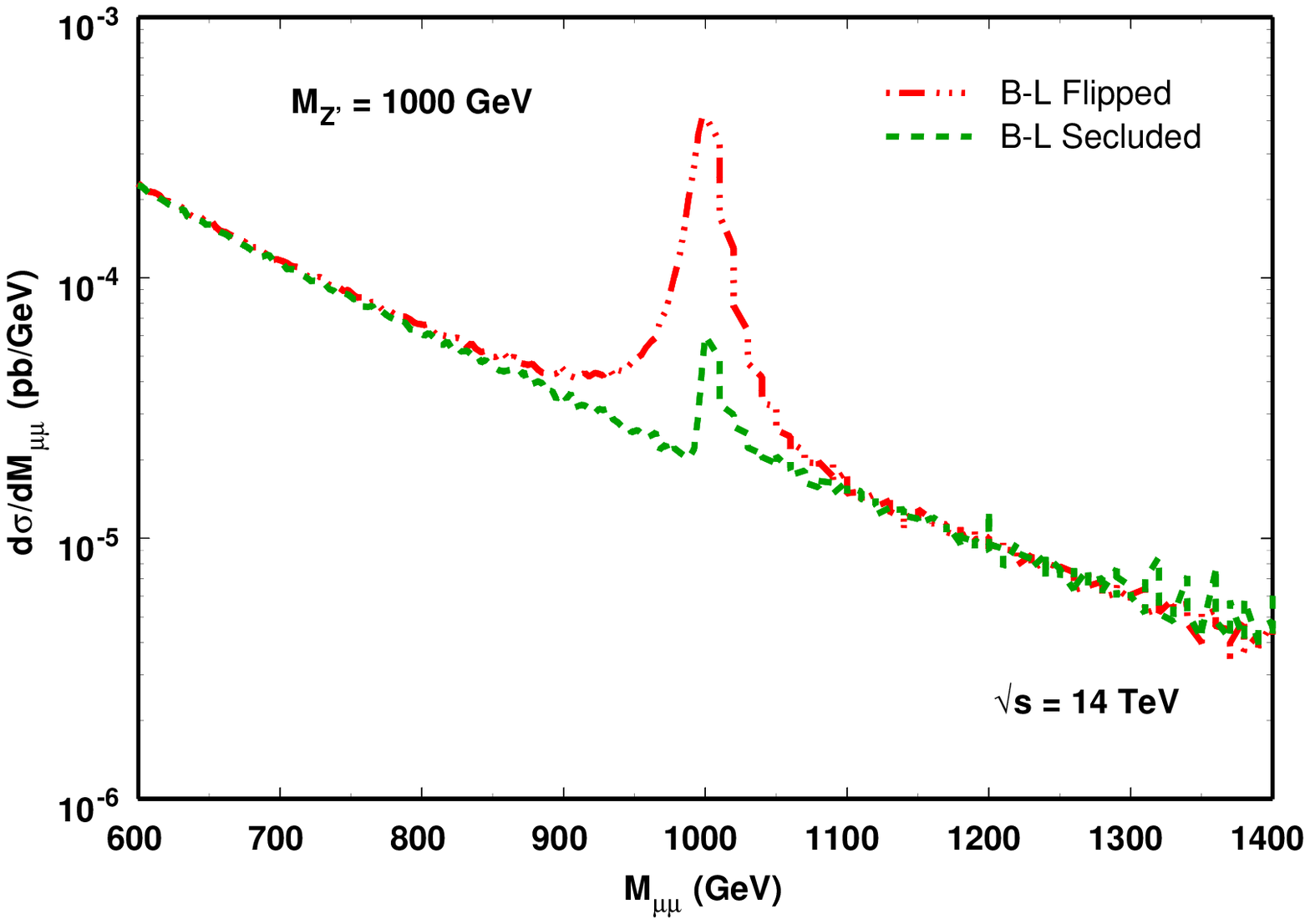}
\end{center}
\caption{ The dilepton invariant mass distribution for $ p
+ p \to \mu^{+} + \mu^{-} + X$ process in Flipped and Secluded models considering $M_{Z^{\prime}}=1000$ GeV at  $\sqrt s = 7$ TeV (up) and $\sqrt s = 14$ TeV (down) with the first cuts set.}
\end{figure}

\begin{figure}[ht]
\begin{center}
\includegraphics[height=.3\textheight]{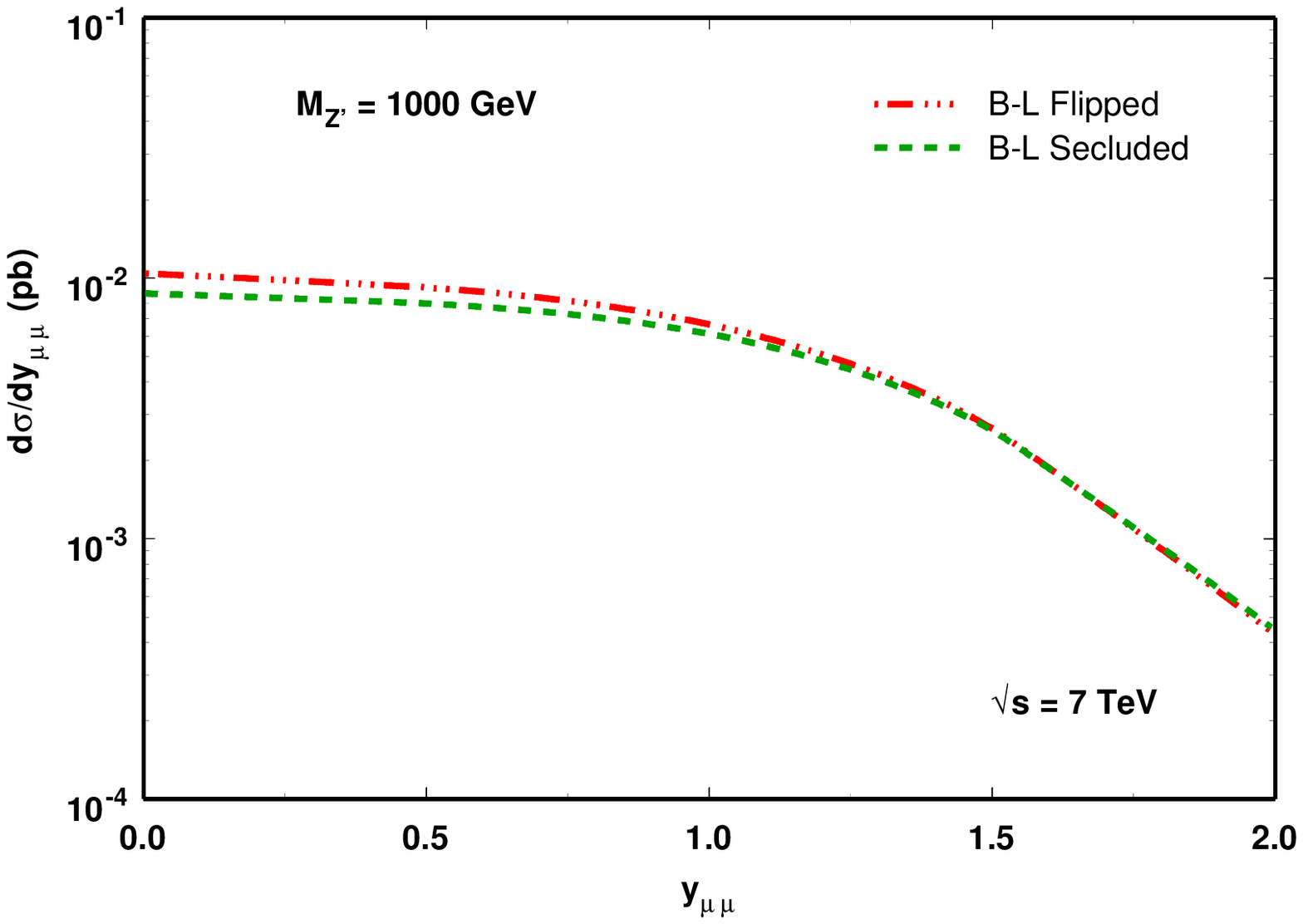}
\includegraphics[height=.3\textheight]{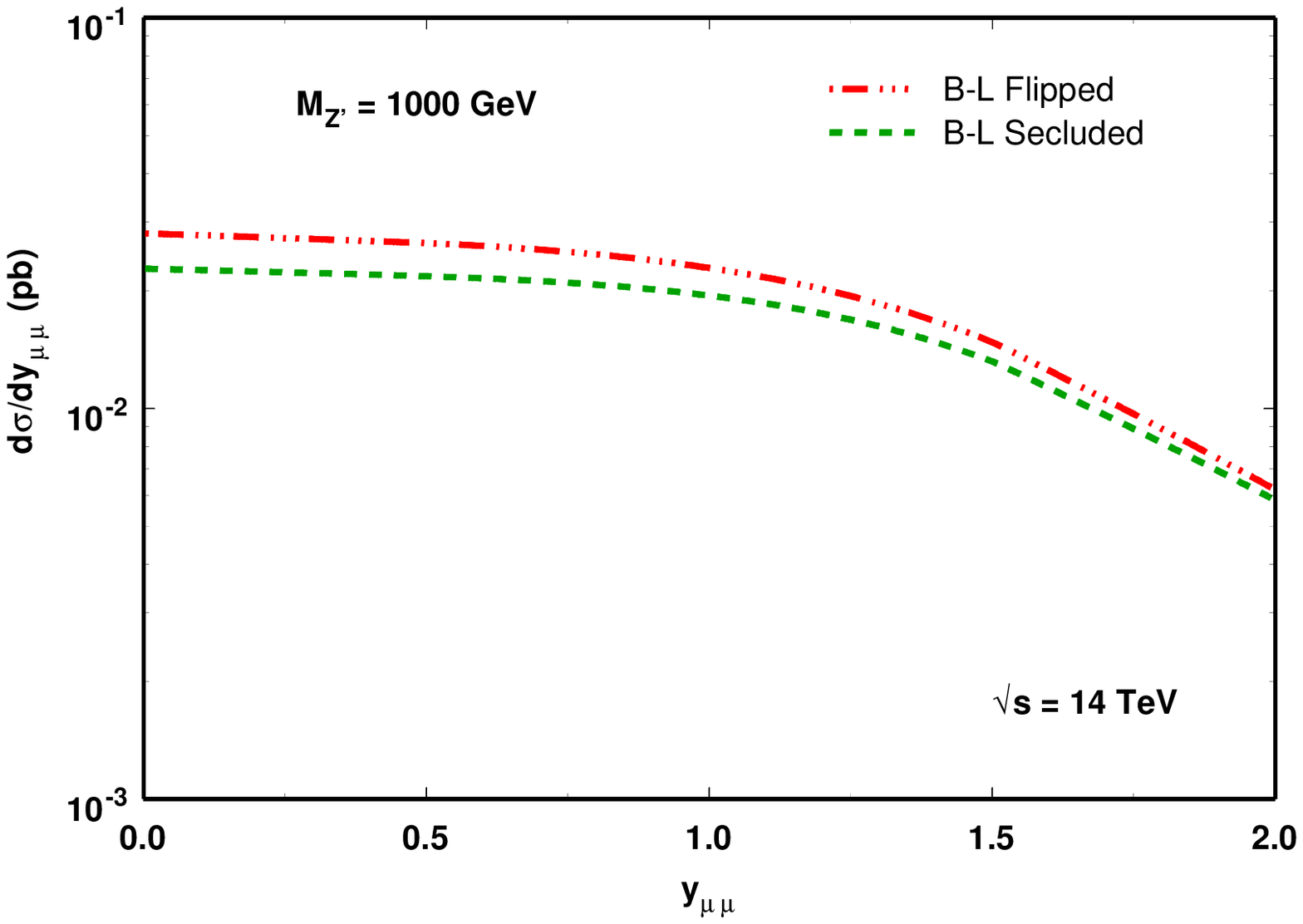}
\end{center}
\caption{ The rapidity distribution for the process $ p + p \to \mu^{+}
+ \mu^{-} + X$  for both $B-L$  models, considering $M_{Z^{\prime}}= 1000$ GeV, $\sqrt s = 7$ TeV (up) and $\sqrt s = 14$ TeV (down) with the first cut set. }
\end{figure}

\begin{figure}[ht]
\begin{center}
\includegraphics[height=.3\textheight]{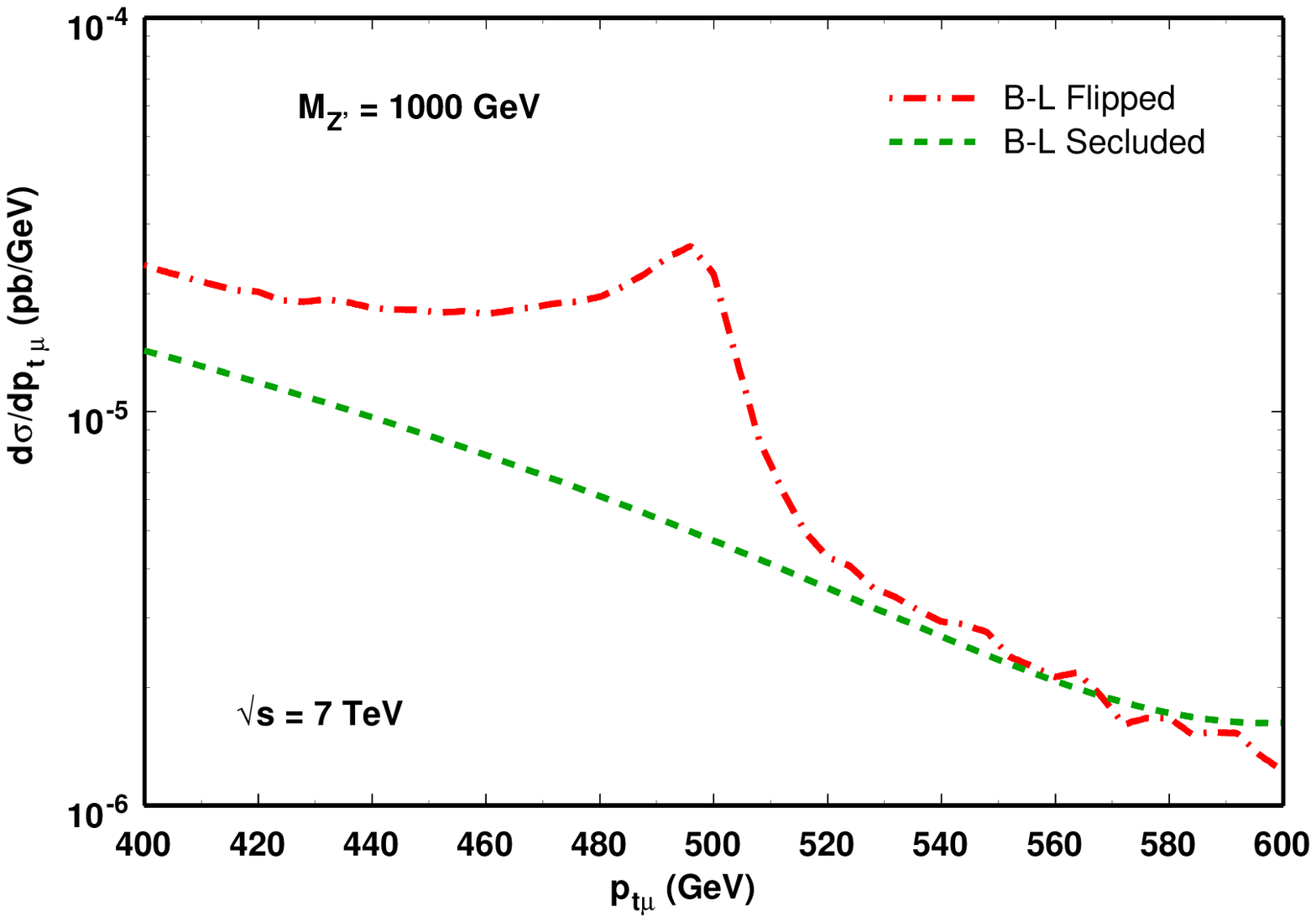}
\includegraphics[height=.3\textheight]{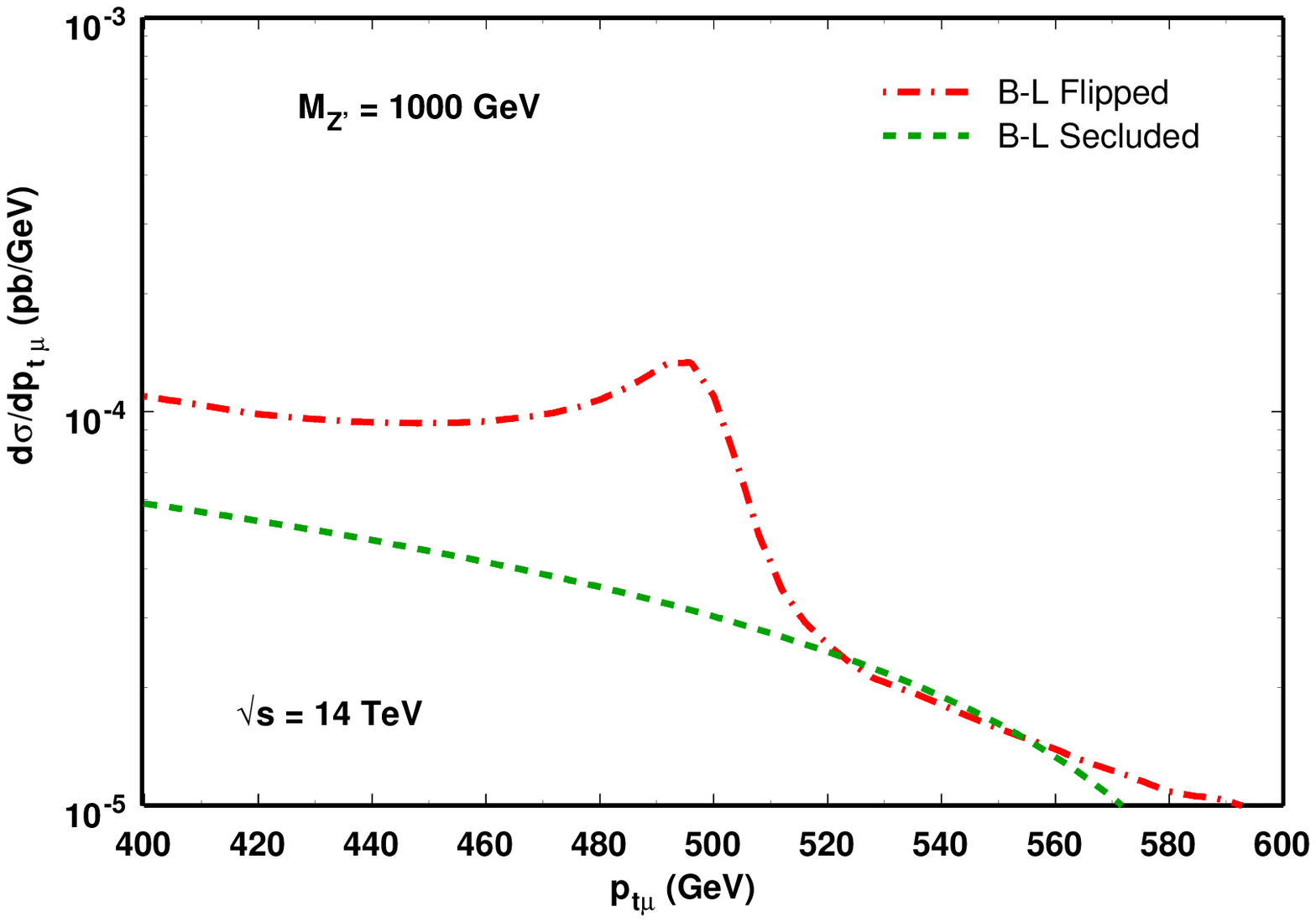}
\end{center}
\caption{ The muon transverse momentum distribution for the process
$ p + p \to \mu^{+} + \mu^{-} + X$ for both $B-L$  models, considering $M_{Z^{\prime}}= 1000$, $\sqrt s = 7$ TeV (up) and $\sqrt s = 14$ TeV (down), applying the first cut set. }
\end{figure}

\begin{figure}[ht]
\begin{center}
\includegraphics[height=.3\textheight]{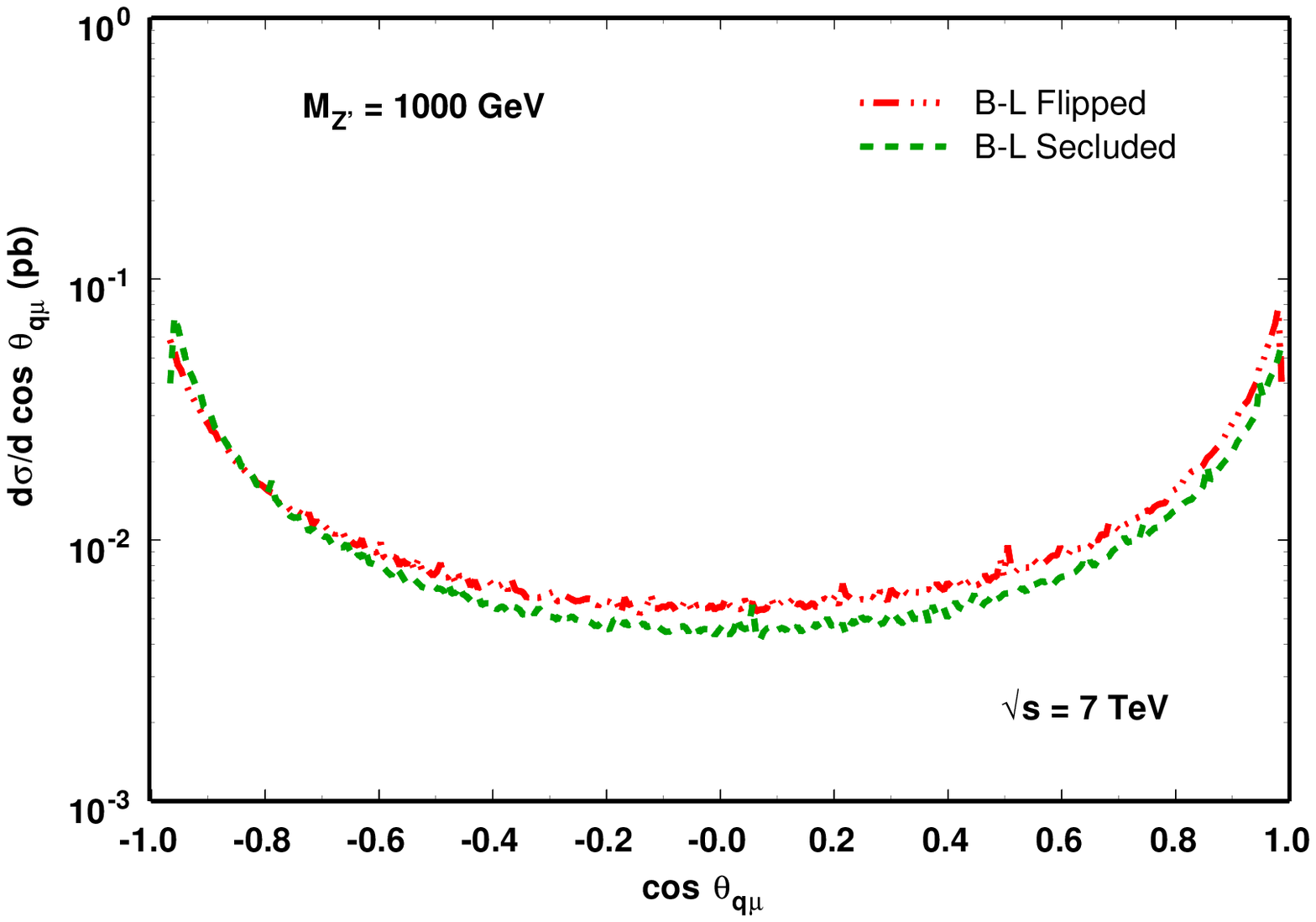}
\includegraphics[height=.3\textheight]{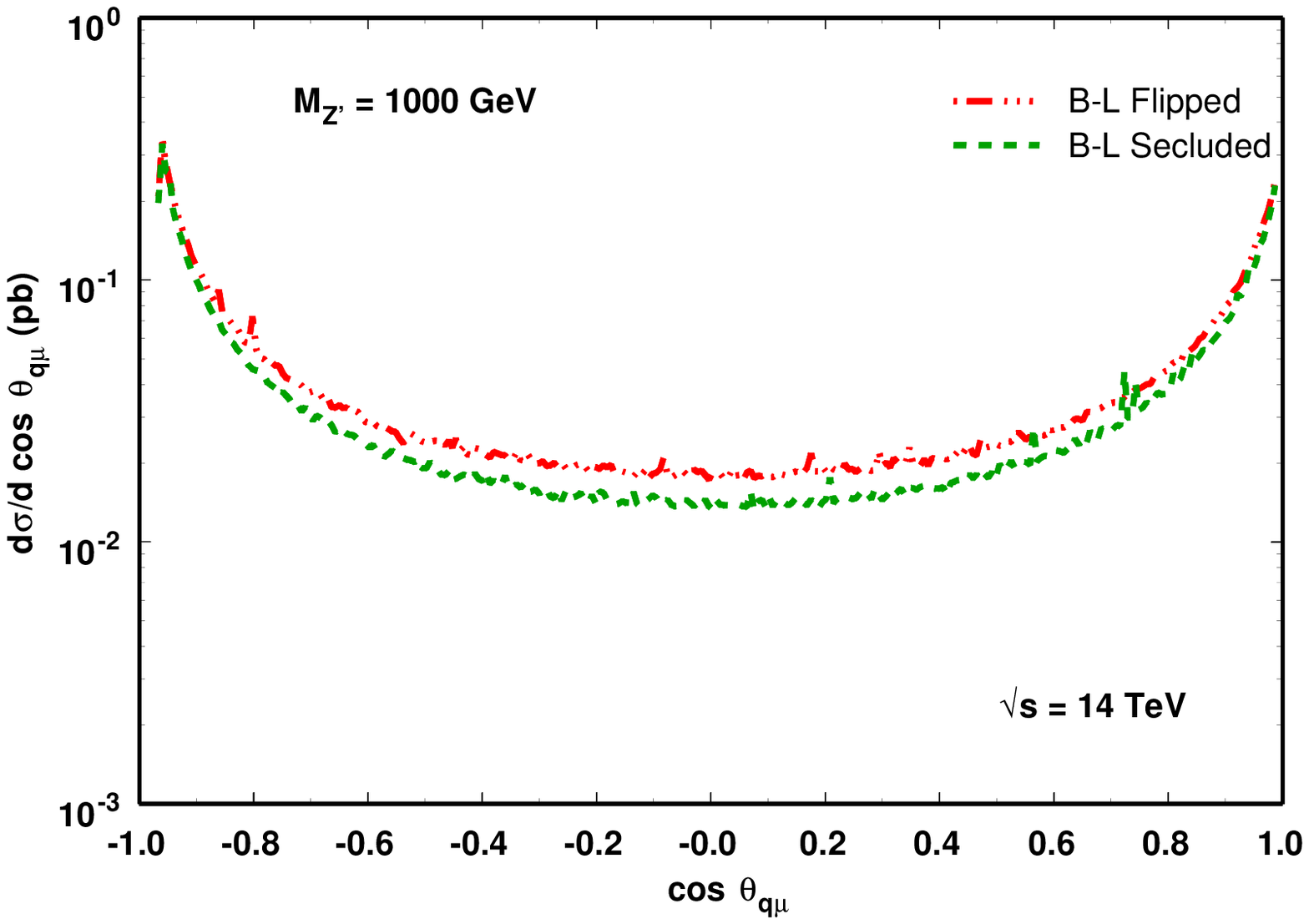}
\end{center}
\caption{ The muon angular distribution for the process
$ p + p \to \mu^{+} + \mu^{-} + X$ for both $B-L$  models, considering $M_{Z^{\prime}}= 1000$, $\sqrt s = 7$ TeV (up) and $\sqrt s = 14$ TeV (down), applying the first cut set. }
\end{figure}

\section{Conclusions}

In order to explore the properties of a new neutral vector boson belonging to
models with $B-L$ local symmetry, we study the Drell-Yan channel with a muon pair production,
for two LHC proposed scenarios: $\sqrt{s}=7$ TeV  (${\cal L} = 1$ fb$^{-1}$) and $\sqrt{s}=14$ TeV (${\cal L} = 100$ fb$^{-1}$). Our study  is focused on two values of the $Z^{\prime}$ boson mass, $1000$ GeV and $1500$ GeV.
\\

First we can say that if $M_{Z^{\prime}}$ is
smaller than the centre of mass energy of the colliding protons,
it can be discovered at the LHC. For instance, for $M_{Z^\prime} = 1500$ GeV and  $\sqrt{s}=14$ TeV we obtain
$\sim 10^8$ events for both $B-L$ models. The signals of $Z^{\prime}$ bosons come from their direct production.

In order to perform $Z^\prime$ studies in $pp$ collisions, many
observables can be used. We can cite the total cross-section and
also several asymmetries very well studied by  $e^{+} e^{-}$ colliders.
The measurement of some of these observables is very difficult due to the
signal, which needs to be detected above the background of the
hadronic experiment. Thus, it is necessary to apply cuts taking into account the
detector acceptance for reducing the background contribution.

The invariant mass distribution of the muon pair final state  is centered
around the $Z^\prime$ mass. This behavior allows us to separate
the signal from the background and to measure the
$Z^\prime$ boson mass. Several cuts can be applied to guarantee more
precision on  measurements. The signal from the Flipped model is above the one from the Secluded model, and the corresponding 
difference can be enhanced if  more severe cuts are considered. On the other hand, the SM background remains below the signals
of both $B-L$ models, as shown in Figure 2.

The muon transverse momentum distribution centered around the half of $Z^\prime$ boson mass is a helpful observable for 
distinguishing the models. In this case we have observed a dominance of the Flipped model over the Secluded one, which  has a 
behavior similar to the SM. See Figure 5.

The forward-backward charge asymmetry is another powerful tool to
provide some information about the $Z^\prime$ couplings to quarks
in $pp$ collisions, when  leptons are in the final state, even being
measured indirectly due to the unknown of original quark
direction. To guarantee the correct direction of the initial parton we applied a cut on final muon pair 
$\vert y_{\mu\mu} \vert  >0.8$.  The resulting $A_{FB}$ asymmetries, for both B-L models, present  clear different behaviors, 
mainly for the muon pair invariant mass in the range $1300-1500$ GeV, as shown in Figure 4.

In addition to $A_{FB}$, the rapidity  distributions are very sensitive to the $Z^\prime$
couplings to quarks and in their computing we can consider  light
and  heavy quarks. They are very useful to disentangle the different
models. Different cuts on it should be analyzed to exploit
different responses from the same experiment. The combination of the   forward-backward asymmetry and rapidity distribution generates another interesting observable, $A_{FBy1}$. If the LHC discovers a  $Z^\prime$ boson, the $A_{FBy1}$  can be used as a refinement  to select to which model the $Z^\prime$ boson belongs to. This combined observable in not under consideration in this work.

Besides the study for LHC operating in the high energy regime, we have performed an analysis at $\sqrt{s}=7$ TeV for
$M_{Z^{\prime}}= 1000$ GeV. The obtained results present a clear signature for the  $Z^\prime$ boson production. The dimuon 
invariant mass distribution shows a sharp peak around $Z^\prime$ boson mass, above the SM background, as shown in Figure 7. 
We note that the transverse momentum distribution once again reveals a peak on $p_T=M_{Z^{\prime}}/2$. For these two 
distributions, the ones of the Flipped model are far  above those of the Secluded model.

In general, the presented models can be well exploited by using the
cited observables above at the LHC  first energy stage. Some of them are more suitable for
distinguishing the models. The choice of the applied cuts is also important. For instance,  the same cuts that we had used to 
the forward-backward asymmetries are not appropriate to the rapidity distributions. Therefore, a combined analysis and
different cuts should be applied to disentangle the models and
guarantee more realist theoretical prediction on the $Z^{\prime}$
couplings to fermions. The LHC running at $\sqrt s = 7$ TeV already is a powerful tool for unraveling  physics beyond the SM.

\vskip 1.cm
\nl
Acknowledgements E. C. F. S. Fortes was supported by FAPESP under
the process 07/59398-2; J. C. Montero was partially supported by
CNPq under the process 307807/2006-1; Y. A. Coutinho thanks FAPERJ
for financial suport.


\begin{thebibliography}{99}
\bibitem{Bar} D.~P.~Barber {\it et al.}, Phys. Rev. Lett. \textbf{43}, 830 (1979).
\bibitem{Lhcilc} G.~Weiglein \textit {et al.} (LHC/LC Study Group), Phys. Rept. \textbf{426}, 47 (2006).
\bibitem{lrgut} R.~W.~Robinett and J.~L.~Rosner, Phys. Rev.  D \textbf{25},
(1985) 3036; P.~Langacker and M.~Luo, \textit{ibid}. D \textbf{45},
(1992) 278; J.~Erler and P.~Langacker, Phys. Rev. Lett. \textbf{84}, (2000) 212.
\bibitem{e6}  D.~London and J.~L.~Rosner, Phys. Rev. D \textbf{34}, (1986) 1530;
J.~Kang and P.~Langacker, Phys. Rev. D \textbf{71} (2005) 035014; P.~Langacker, R.~W.~Robinett and J.~L.~Rosner, Phys. Rev. D \textbf{30}, 1470 (1984); J.~L.~Hewett, T.~G.~Rizzo, Phys. Rep. \textbf{183}, 193 (1989).
\bibitem{lr} A.~Davidson, Phys. Rev. D \textbf{20}, (1979) 776; V.~Barger, E.~Ma
and K.~Whisnant, Phys. Rev. D \textbf{26}, (1982) 2378; M.~K.~
Parida and A.~Raychaudhuri, Phys. Rev. D \textbf{26}, (1982) 2364.
\bibitem{Lit} N.~Arkani-Hamed, A.~G.~Cohen and H.~Georgi, Phys. Lett. B {\bf 513}, 232 (2001) ; N.~Arkani-Hamed, A.~G.~Cohen, E.~Katz and A.~E.~Nelson, JHEP {\bf 0207}, 034 (2002).
\bibitem{Ssm} Vernon D. Barger, Wai-Yee Keung, Ernest Ma. Phys. Lett. B \textbf{94},(1980) 377;
Vernon D. Barger, Wai-Yee Keung, Ernest Ma. Phys. Rev. D \textbf{22} (1980) 727;
Vernon D. Barger, Wai-Yee Keung, Ernest Ma. Phys. Rev. Lett. \textbf{44} (1980) 1169;
H. Georgi and S. Weinberg, Phys. Rev. D \textbf{17} (1978) 275;  E. H. De Groot, D. Schildknecht and G. J. Gounaris, University of Bielefeld Report No. BI-TP 79/37 (unpublished).
\bibitem{Riz} ``Pedagogical Introduction to Extra Dimensions", Thomas G.~Rizzo, SLAC-PUB-10753,  SSI-2004-L013, Sep 2004, arXiv:hep-ph/0409309v2.
\bibitem{PIV} F.~Pisano and V.~Pleitez, Phys. Rev. D {\bf 46}, 410 (1992).
\bibitem{FRA} P.~H.~Frampton, Phys. Rev. Lett. {\bf 69}, 2889 (1992).
\bibitem{TON} J.~E.~Cieza Montalvo, M.~D.~Tonasse, Phys. Rev. D {\bf67}, 075022 (2003).
\bibitem{Bess} De Curtis, Stefania, The BESS model revisited as a Higgsless LinearMoose $@$ the LHC,  arXiv:1002.2361.
\bibitem{Appel} T.~Appelquist, B.~A.~Dobrescu, and A.~R.~Hopper, Phys. Rev. D \textbf{68}, 035012 (2003).
\bibitem{Blsm} J.~C.~Montero and V.~Pleitez, Phys. Lett. B \textbf{765}, 64 (2009).
\bibitem{Elai} E.~C.~F.~S.~Fortes, J.~C.~Montero, V.~Pleitez,  Phys. Rev. D \textbf{82}, 114007 (2010).
\bibitem{Khalil} W. Emam, and S. Khalil, Eur. Phys. J. C52, 625 (2007).
\bibitem{Abu} A.~Abulencia {\it et al.}, (CDF Collaboration), Phys. Rev. Lett. 96, 211801 (2006); T.~Aaltoni {\it et
al.}, (CDF Collaboration), Phys. Rev. Lett. \textbf{102}, 091805 (2009).
\bibitem{Erler} J.~Erler, P.~Langacker, S.~Munir, and E.~Rojas, JHEP \textbf{08}, 017 (2009).
\bibitem{Caccia} G.~Cacciapaglia, C.~Cs\'aki, G.~Marandella, and A.~Strumia, Phys. Rev. D \textbf{74}, 033011 (2006).
\bibitem{Salvioni} E.~Salvioni, G.~Villadoro, and F.~Zwirner, JHEP \textbf{11}, 068 (2009).x
\bibitem{Pdg} K.~Nakamura {\it et al.} (Particle Data Group), J. Phys. G \textbf {37}, 075021 (2010).
\bibitem{Pam} M.~Boezio {\it et al.}, New J. Phys. \textbf{11}, 105023 (2009).
\bibitem{Comp} E.~Boos {\it et al.}, [CompHEP Collaboration], CompHEP 4.4: Automatic computations from Lagrangians to events, Nucl. Instrum. Meth. A \textbf{534}, 250 (2004).
\bibitem{Dit} Michael Dittmar, Phys. Rev. D \textbf{55}, 161 (1997).
\bibitem{Nic} Michael Dittmar, Anne-Sylvie Nicollerat, Abdelhak Djouadi, Phys. Lett. B \textbf{583}, 111 (2004).
\bibitem{Elm} E.~Ramirez Barreto, Y.~A.~ Coutinho, J.~S\'a Borges,  Phys. Lett. B. \textbf{689}, 36 (2010).
\end{thebibliography}
\end{document}